\documentclass{webofc}
\usepackage[english]{babel}
\usepackage[utf8]{inputenc}
\usepackage{amsmath,ulem}
\usepackage{amssymb}
\usepackage{graphicx}
\usepackage{color}
\usepackage{float}
\usepackage{slashed}
\usepackage{lipsum}
\usepackage[varg]{txfonts}   

\definecolor{green}{RGB}{0,200,0}

\newcommand\numberthis{\addtocounter{equation}{1}\tag{\theequation}}

\begin{document}
\title{Hybrid stars from a constrained equation of state}

        \author{\firstname{Márcio Ferreira}\inst{1}\fnsep\thanks{\email{marcio.ferreira@uc.pt}} 
        \and \firstname{Renan Câmara Pereira} \inst{1}\fnsep\thanks{\email{renan.pereira@student.uc.pt}}
        \and \firstname{Constança Providência} \inst{1}\fnsep\thanks{\email{cp@fis.uc.pt}}}

\institute{CFisUC, 
	Department of Physics, University of Coimbra, P-3004 - 516  Coimbra, Portugal}

\abstract{%
We determine, within a meta-model, the properties of the nuclear matter equation of state (EoS) that allow for a phase transition to deconfinement matter. It is shown that the properties that define the isoscalar channel are the ones that are affected, in particular,  a phase transition implies much larger values of the skewness and kurtosis. 
The effect of multi-quark interaction channels in the description of the
quark phase in hybrid stars is also studied.  NS properties, such as the mass and radius of the quark core,
show an interplay dependence between the 8-quark vector and the 4-quark
isovector-vector interactions. We show that low mass NS, $M\sim 1.4 M_\odot$,
may already contain a quark core, and satisfy all  existing NS observational constraints. We discuss the strangeness content of the quark core and its influence on the speed of sound.
}
\maketitle
\section{Introduction}
\label{intro}

The physics of neutron stars (NS) can be studied by combining astrophysical observations of
electromagnetic radiation, astro-particles and gravitational waves (GW) -- multi-messenger astrophysics.
The recent observation of the massive pulsars, PSR J0348+0432 with $M=2.01\pm$0.04 $M_\odot$
\cite{Antoniadis:2013pzd} and 
or MSP J0740$+$6620,
 a mass 
$2.08^{+0.07}_{-0.07}M_\odot$ 
\cite{Cromartie2019,Fonseca2021}
 impose strong constraints on the equation of state (EoS) of nuclear matter.
The estimation of an upper bond on tidal deformability of a NS star by the LIGO/Virgo collaboration from the GW170817 event 
is important in further constraining the high density behavior of the NS EoS \cite{TheLIGOScientific:2017qsa,LIGOScientific:2018cki}.
While a not too stiff EoS is favored from the upper bond on the tidal deformability, massive pulsars require, on the other hand, a considerably stiff EoS for the NS. Such apparent tension might indicate the presence of exotic matter in NS, i.e., non-nucleonic degrees of freedom.  The transition, via a first-order phase transition, from hadronic to quark matter at high densities could be an explanation for the existence of low tidal deformabilities for NS masses around $1.4M_{\odot}$. 
There is a great interest in finding observational signatures sensible to 
the presence of exotic matter inside NS \cite{Most:2018eaw,Alford:2019oge,Weih:2019xvw}.

Herein, hybrid NS are described by applying a two-model approach: we choose an hadronic model
for the confined phase and a quark model for the quark phase, which are connected through a first-order phase transition (applying the Maxwell construction). The quark model chosen is the well known Nambu-Jona--Lasinio (NJL) model, 
an effective model of QCD, that considers chiral symmetry preserving interactions \cite{Hatsuda:1994pi,Buballa:2003qv}. 
The use of the NJL model in such two-model approaches has been widely explored both considering
a first-order phase transition \cite{Pagliara:2007ph,Benic:2014iaa,Benic:2014jia,Zacchi:2015oma,CamaraPereira:2016chj,Wu:2018kww} and with a smooth (crossover) transition between both phases \cite{Baym:2017whm,Kojo:2018glx}. Starting from the three-flavor NJL model, we study the effect of several interaction channels 
in the existence and stability of hybrid stars, and specifically on quark core
properties \cite{Ferreira:2020evu,Ferreira:2020kvu,Ferreira:2021osk}.

\section{Equation of state of hybrid NS}
\label{eos}
The matter inside hybrid stars is modelled via a two-model approach: an hadronic phase
is connected to a quark phase  through a first-order phase transition, i.e., a Maxwell construction is implemented. Therefore, we need to choose both an hadronic model and a quark model.

\subsection{Hadronic EoS}

In the following, we consider two types of hadronic EoS:   a) a set of metamodels  based on a Taylor expansion around saturation density \cite{margueron2018}; b) a relativistic mean-field model with density dependent couplings  DDME2 \cite{ddme2}, which satisfies well  accepted nuclear matter properties at saturation ($K_{sat}=250$~MeV, $E_{sym}=32.3$~MeV,    $L_{sym}=51.2$~MeV, $K_{sym}= -87$~MeV) , and besides describes two solar mass stars.

In the following, we summarize the metamodel considered.  
The energy per particle functional of homogeneous nuclear matter, $\mathcal{E}$, can be written as:
\begin{align}
\mathcal{E} (n_n,n_p) = 
e_{sat}(n) +
e_{sym}(n)\delta^2 .
\end{align}
where  $n_n$ and $n_p$ are  neutron and proton densities, $n=n_n+n_p$ is the baryonic density and $\delta=(n_n-n_p)/n$ is the asymmetry. 
Following previous studies, e.g.  \cite{margueron2018}, we parametrize
the hadronic EoS as a Taylor expansion around the saturation density. We consider a fourth order expansion around saturation density, $n_{0}$:
\begin{align}
e_{sat}(x) & = \sum_{n=0}^4 \frac{1}{n!} P_{IS}^{(n)} x^n , \qquad  e_{sym}(x)=\sum_{n=0}^4 \frac{1}{n!} P_{IV}^{(n)}x^n 
\end{align}
where, $x=\frac{(n-n_{0})}{3n_{0}}$.  The coefficients of $e_{sat}(n)$ and  $e_{sym}(n)$ can be identified, respectively,  with the isoscalar and isovector empirical parameters,
\begin{align}
P_{IS}^{(k)} = 
(3n_{0})^k\left.\frac{\partial^k e_{sat}}{\partial n^k}\right|_{\{\delta=0,n=n_{0}\}} , \qquad
P_{IV}^{(k)}= (3n_{0})^k\left.\frac{\partial^k e_{sym}}{\partial n^k}\right|_{\{\delta=0,n=n_{0}\}}.
\end{align}
The correspondence between the coefficients and the empirical parameters can then be written as:
${E_{sat},K_{sat},  Q_{sat},Z_{sat} } \rightarrow {P_{IS}^{(0)},P_{IS}^{(2)},P_{IS}^{(3)},P_{IS}^{(4)} }$, and
${E_{sym},L_{sym}, K_{sym},  Q_{sym},Z_{sym} } \rightarrow {P_{IV}^{(0)},P_{IV}^{(1)},P_{IV}^{(2)},P_{IV}^{(3)},P_{IV}^{(4)}}.
$

The  advantages of taking a  meta-model for the hadron phase can be summarized in the  following two points: (i) its flexibility to build an hybrid EoS with 1.97$M_\odot$ maximum mass; (ii) the possibility of choosing the empirical nuclear parameters that
satisfy uncertainties on the EoS parameters known from experiments and other theoretical studies. From all the meta-model EoSs built, 354 of them  allow for a phase transition to quark matter.   In Table \ref{tab1}, the mean and standard deviation of the properties of the both sets of EoS,  nucleonic and hybrid, are given.  We also include a third set of EoS that allows for a phase transition to quark matter at a quite low baryonic density, i.e. the transition density $n_t$ satisfies $1.3<n_t/n_0<2.5$.
 Comparing the three meta-models sets we conclude that  a phase transition  to quark matter requires    
  larger values of isoscalar properties, in particular, $Q_{sat}$ and $Z_{sat}$, and a  wider spread, while the isovector channel is essentially not affected.

\begin{table}[h]
    \caption{The mean $\overline{P}_{i}$ and standard deviation
$\sqrt{\sigma_{{P}_{i}}}$ of the multivariate Gaussian, where
$\sigma_{{P}_{i}}$ is the variance of the parameter $P_{i}$. 
Our EoSs are sampled using the initial distribution for 
$P_i$ assuming that there are no correlations
among the parameters. All the quantities are in units of MeV.
The values of $E_{sat}$ and $n_{0}$ are fixed to $-15.8$ MeV and $0.155$ fm$^{-3}$,
respectively. The parameters are given for three scenarios always considering that a two solar mass star is described by the models: (i) purely nucleonic EoS; (ii) hybrid EoS, all without furhter filters; (iii) hybrid stars with a phase transition in the range $1.3n_0 - 2.5n_0$. }
\small{
       \begin{tabular}{ccccccccc}
          \hline
         {(i) Nucleonic}    	& $K_{sat}$ & $Q_{sat}$ & $Z_{sat}$ &
                                                                              $E_{sym}$ & $L_{sym}$ & $K_{sym}$ & $Q_{sym}$ & $Z_{sym}$ \\
         \hline
         mean&  {233.95}&  58.62 &  -181.97 &  33.32 & 51.56  & -43.96  & 238.21 & 371.91\\
         std& 18.75 &123.33 &143.04 & 1.89 & 11.83 & 63.02 & 300.33 & 698.56\\
         \hline       
         {  (ii) Hybrid (all)}		& $K_{sat}$ & $Q_{sat}$ & $Z_{sat}$ & $E_{sym}$ & $L_{sym}$ & $K_{sym}$ & $Q_{sym}$ & $Z_{sym}$ \\ 
         \hline
         mean & { 236.52} & { 241.38} & { 362.11} & 33.17 & 49.99 & -37.48 & 191.54 & 503.39 \\ 
         std & 17.73 & 242.03 & 593.75 & 1.84 & 12.55 & 67.23 & 310.48 & 734.69
         \\
        \hline
         
         {(iii) Hybrid }		& $K_{sat}$ & $Q_{sat}$ & $Z_{sat}$ & $E_{sym}$ & $L_{sym}$ & $K_{sym}$ & $Q_{sym}$ & $Z_{sym}$ \\ 
         ({\small $1.3<n_t/n_0<2.5$)}&\\
         \hline
         mean &  {244.55} & { 434.37} & { 557.83} & 33.38 & 49.90 & -18.07 & 195.39 & 400.46 \\ 
         std & 16.91 & 174.6 & 743.94 & 2.00 & 11.55 & 57.47 & 300.13 & 672.24
         \\
                  \hline
       \end{tabular}
}       

    \label{tab1}
\end{table}

To build an hadronic EoS, we use random sampling to choose a point in the 8-dimensional space of parameters from a multivariate Gaussian with zero covariance:
\begin{equation}
\text{EoS}_i = (E_{sym},L_{sym},K_{sat},K_{sym},Q_{sat},Q_{sym},Z_{sat},Z_{sym})_i \,\, \sim N(\boldsymbol{\mu},\boldsymbol{\Sigma})
\label{eq:hadronic1}
\end{equation}
with the mean defined by 
$
\boldsymbol{\mu}^T=(\overline{E}_{sym},\overline{L}_{sym},\overline{K}_{sat},\overline{K}_{sym},\overline{Q}_{sat},\overline{Q}_{sym},\overline{Z}_{sat},\overline{Z}_{sym}),
$
and the  covariance matrix given by
$
\boldsymbol{\Sigma}=diag(\sigma_{E_{sym}},\sigma_{L_{sym}},\sigma_{K_{sat}},\sigma_{K_{sym}},
\sigma_{Q_{sat}}, \sigma_{Q_{sym}},\sigma_{Z_{sat}},\sigma_{Z_{sym}}).
$

\subsection{Quark EoS}
For the quark matter phase, we use a SU$(3)$ NJL-type model, given by the following multi-quark interaction Lagrangian density:
\begin{align*}
\mathcal{L} &=  
\bar{\psi} 
(
i\slashed{\partial} - \hat{m} + \hat{\mu} \gamma^0 
) 
\psi 
 + G_S  \sum_{a=0}^8
[ (\bar{\psi} \lambda^a \psi)^2 + 
(\bar{\psi} i \gamma^5 \lambda^a \psi)^2 ]
 - G_D [  
\det( \bar{\psi} (1+\gamma_5) \psi ) + 
\det( \bar{\psi} (1-\gamma_5) \psi )  ]
\\
& - G_\omega [ (\bar{\psi}\gamma^\mu\lambda^0\psi)^2 + (\bar{\psi}\gamma^\mu\gamma_5\lambda^0\psi)^2 ]
 - G_\rho \sum_{a=1}^8 
[ 
(\bar{\psi} \gamma^\mu\lambda^a \psi)^2 +  
(\bar{\psi} \gamma^\mu\gamma_5\lambda^a \psi)^2 
]\\
&
- G_{\omega \omega} 
[ (\bar{\psi}\gamma^\mu\lambda^0\psi)^2 + (\bar{\psi}\gamma^\mu\gamma_5\lambda^0\psi)^2 ]^2
\\
& - G_{\rho \rho} \sum_{a=1}^8 \sum_{b=1}^8 
[ 
(\bar{\psi} \gamma^\mu\lambda^a \psi)^2 +  
(\bar{\psi} \gamma^\mu\gamma_5\lambda^a \psi)^2 
]
[ 
(\bar{\psi} \gamma^\mu\lambda^b \psi)^2 +  
(\bar{\psi} \gamma^\mu\gamma_5\lambda^b \psi)^2 
]
\\
& - G_{\omega \rho} 
[ (\bar{\psi}\gamma^\mu\lambda^0\psi)^2 + (\bar{\psi}\gamma^\mu\gamma_5\lambda^0\psi)^2 ]
\sum_{a=1}^8 
[ 
(\bar{\psi} \gamma^\mu\lambda^a \psi)^2 +  
(\bar{\psi} \gamma^\mu\gamma_5\lambda^a \psi)^2 
].\numberthis
\label{eq:SU3_NJL_lagrangian}
\end{align*} 

The  parameters of the quark model, $\hat m=diag(m_u,m_d,m_s)$, $G_S$, $G_D$, and $\Lambda$, are fit to reproduce the vacuum meson masses and decay constants (Table \ref{tab:2}), while the remaining parameters are left as free.

\begin{table}[!htb]
\caption{Parameters of the quark  model: $\Lambda$ is the model cutoff, $m_{u,d,s}$ are the quark current masses, $G_S$ and $G_D$ are coupling constants. $M_{u,d}$ and $M_{s}$ are the resulting constituent quark masses in the vacuum. }
\begin{center}
    \begin{tabular}{cccccccc}
    \hline
    \hline
$\Lambda$  & $m_{u,d}$ & $m_s$  & $G_S\Lambda^2 $ & $G_D\Lambda^5 $ & $M_{u,d}$ & $M_s$ \\
\text{[MeV]}      &  [MeV]    & [MeV]   &               &                  & [MeV] &  [MeV]\\
\hline
   \hline
623.58 & 5.70   & 136.60 & 1.67 &  13.67 & 332.2   & 510.7 \\
   \hline
\end{tabular}
\label{tab:2}
\end{center}
\end{table}

We have analyzed the effect of some of these specific channels on hybrid stars in  \cite{Ferreira:2020evu,Ferreira:2020kvu,Ferreira:2021osk}, where a detailed description can be found. In particular, in the following we discuss the effect of parameters $G_{\omega}$, $G_{\rho}$ and $G_{\omega \omega}$. The parameters $ G_{\rho \rho}$ and $ G_{\omega \rho}$ have only a residual or  small effect on the EoS and NS properties. In the following, we will characterize each quark model by the coupling ratios: $\chi_{\omega} = G_\omega / G_S$, $\chi_{\omega \omega} = G_{\omega \omega} / G_S^4$, $\chi_{\rho} = G_\rho / G_S$, $\chi_{\rho \rho} = G_{\rho \rho} / G_S^4$, and $\chi_{\omega \rho} = G_{\omega \rho} / G_S^4$. 

The thermodynamical potential is determined using the mean-field approximation, from which we calculate the pressure and energy density for cold quark matter, i.e., at zero temperature. The pressure $P$ and energy density $\epsilon$ are defined up to an extra constant term $B$ (the bag constant), i.e.,  $P \to P + B$ and $\epsilon \to \epsilon-B$. This extra parameter will allow us to control the baryonic density at the hadron-quark phase transition.
As referred before, for the hadronic phase, we use the metamodel introduced above and the relativistic mean-field model with density dependent couplings model DDME2 \cite{ddme2}.
The first-order phase transition from hadronic ($H$) to quark ($Q$) matter is accomplished by imposing the Maxwell construction, i.e., $\mu_B^H = \mu_B^Q$ and $P^H = P^Q$, where $\mu_B^{H,Q}$ is the baryon chemical potential of each phase.

\section{Results and discussion}
\label{sec:results}

\begin{figure}[!t]
\begin{center}
\begin{tabular}{c}
\includegraphics[width=0.7\linewidth]{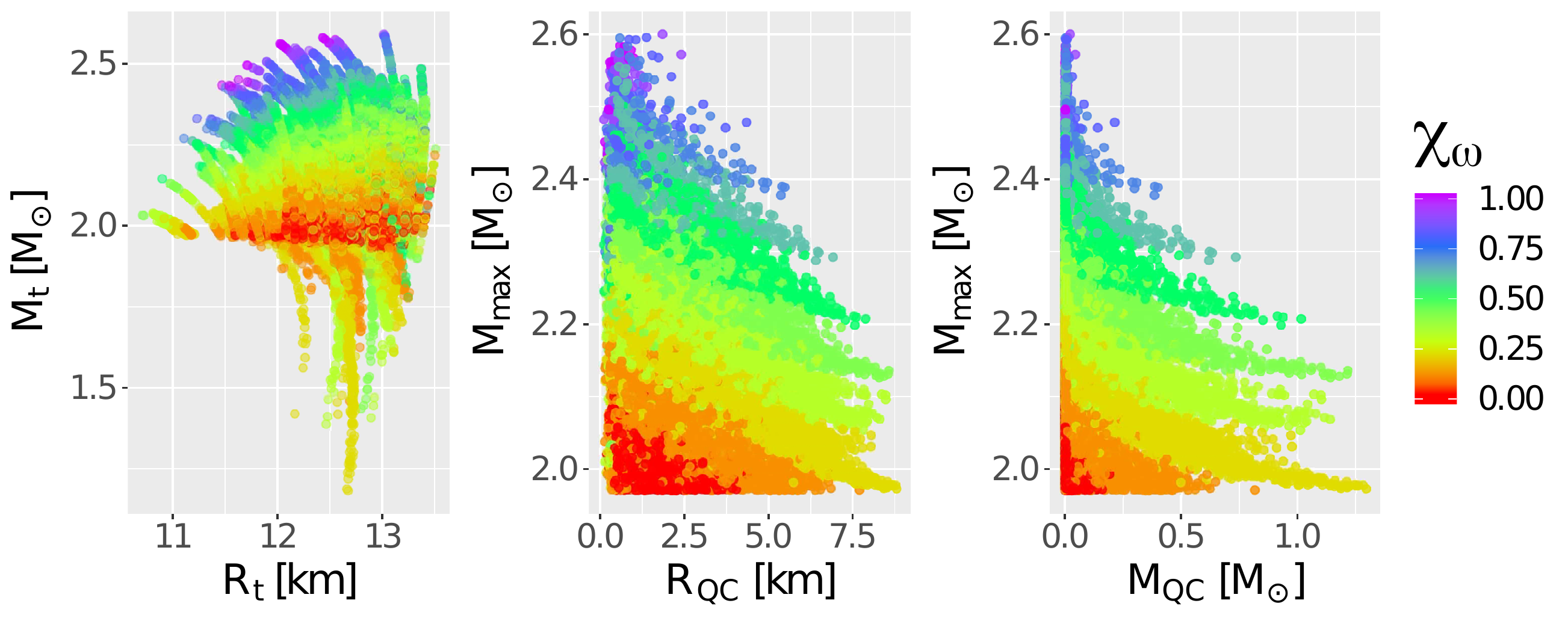}
\end{tabular}
	\caption{The $M_{t}$ vs $R_{t}$ (left), which corresponds to the $(M,R)$ where the onset of quark matter occurs,
		  and the diagrams of maximum NS mass vs. quark core radius (center) and mass (right) as a function of 
		   $\chi_\omega$ (color scale) for each hybrid EoS.}
	\label{fig:qrm}
	\end{center}
\end{figure}

In the following, we study how the quark matter properties affect the existence of stable hybrid stars. First we analize the effect of the vector-isoscalar parameter $\chi_\omega= G_\omega / G_S$ on the values of the mass and radius, respectively, $M_t$ and $R_t$, of the star where quarks start to nucleate  (see Fig. \ref{fig:qrm} left panel) and the radius and mass of the quark core, respectively, $M_{QC}$ and $R_{QC}$, as a function of the maximum NS mass (Fig. \ref{fig:qrm} center and right panels). In this figure each hybrid EoS is represented by one point whose color designates a specific value of $\chi_\omega$, shown in color scale.    The radius and mass of the quark core, $R_{QC}$ and $M_{QC}$, is determined from the quark content of the most massive stable NS
that contains quark matter.
As $\chi_\omega$ increases, the appearance of quarks occurs for larger NS masses and radii.
Quark cores in massive NS, with $M>2.4M_{\odot}$, are attainable with $\chi_\omega>0.6$.
However, the quark core size (center and right panels), of the EoSs that sustain high values of $M_{max}$, is quite small.  The distribution of $M_{QC}$ has a mean value of $0.08M_{\odot}$ and a maximum value of 
$1.29M_{\odot}$, whereas the mean value of $R_{QC}$ is $1.99$ km and the maximum value of $8.78$ km. 
Quark cores having radii as large as 7 km have been also obtained in \cite{annala2019} using a quite different approach.

\begin{figure}[!t]
\begin{center}
\begin{tabular}{c}
\includegraphics[width=0.8\linewidth]{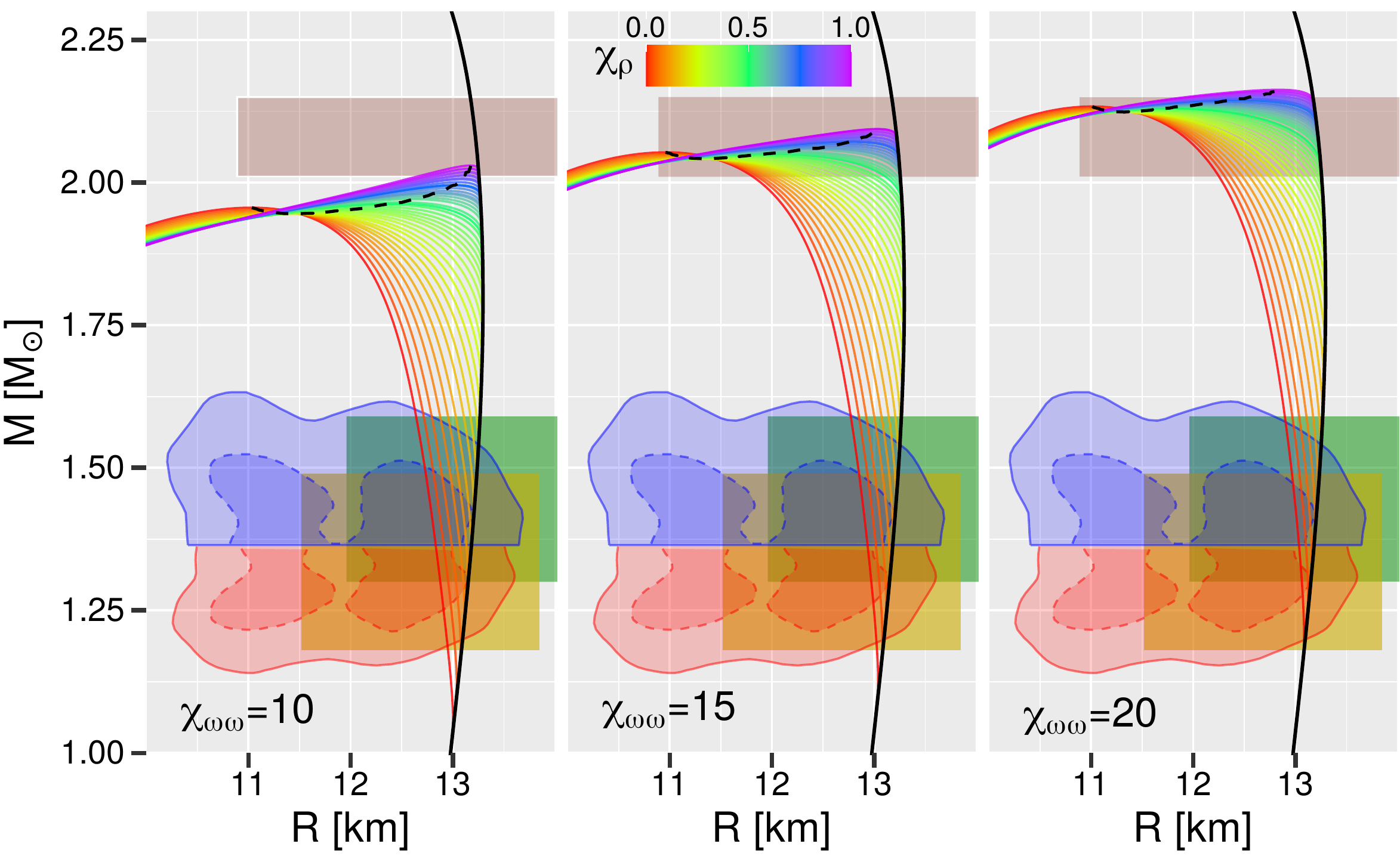}
\end{tabular}
	\caption{$M(R)$ diagrams as a function of $\chi_{\rho}$ (color scale)  for three values of $\chi_{\omega\omega}$: 10 (left), 15 (center), and 20 (right). The solid black line represents the purely hadronic sequence, and the black dashed line indicates the maximum mass of each hybrid EoS. The top blue and bottom red regions indicate, respectively, the 90\% (solid) and 50\% (dashed) credible intervals of the LIGO/Virgo analysis for the GW170817 event \cite{LIGOScientific:2018cki}. The rectangular regions enclosed by dotted lines indicate the constraints from the millisecond pulsar PSR J0030+0451 NICER x-ray data  \cite{Riley:2019yda,Miller:2019cac}. The top brown region is from \cite{Miller2021}.}
	\label{fig:esym2}
	\end{center}
\end{figure}

We next  analyze the impact of $\{\chi_{\omega \omega},\chi_{\rho} \}$ (with $\chi_{\omega}=0$) in the $M(R)$ diagram (Fig. \ref{fig:esym2}).
Larger quark branches are realized for larger $\chi_{\omega \omega}$ values,
where massive NS ($M>2M_\odot$) are already attained for $\chi_{\omega\omega}=15$ and $20$. This is explained because the magnitude of the $\chi_{\omega\omega}$ term is still small at the hadron-quark transition, not affecting the transition density. However, at larger densities it contributes with a strong repulsive term, gives rise to a hard quark EoS, and larger star masses are reached.

The NS mass of the quark onset increases with $\chi_{\rho}$ (regardless of $\chi_{\omega\omega}$)
because the coupling shifts the onset of strange quarks to lower densities, making the quark EoS softer. A second effect of the softening of the EoS is  the decrease of the
 quark branch: the  smoothing of the EoS caused by the appearance of strange quarks prevents too massive stars to occur.

\begin{figure}[!t]
\begin{center}
\begin{tabular}{c}
\includegraphics[width=0.8\linewidth]{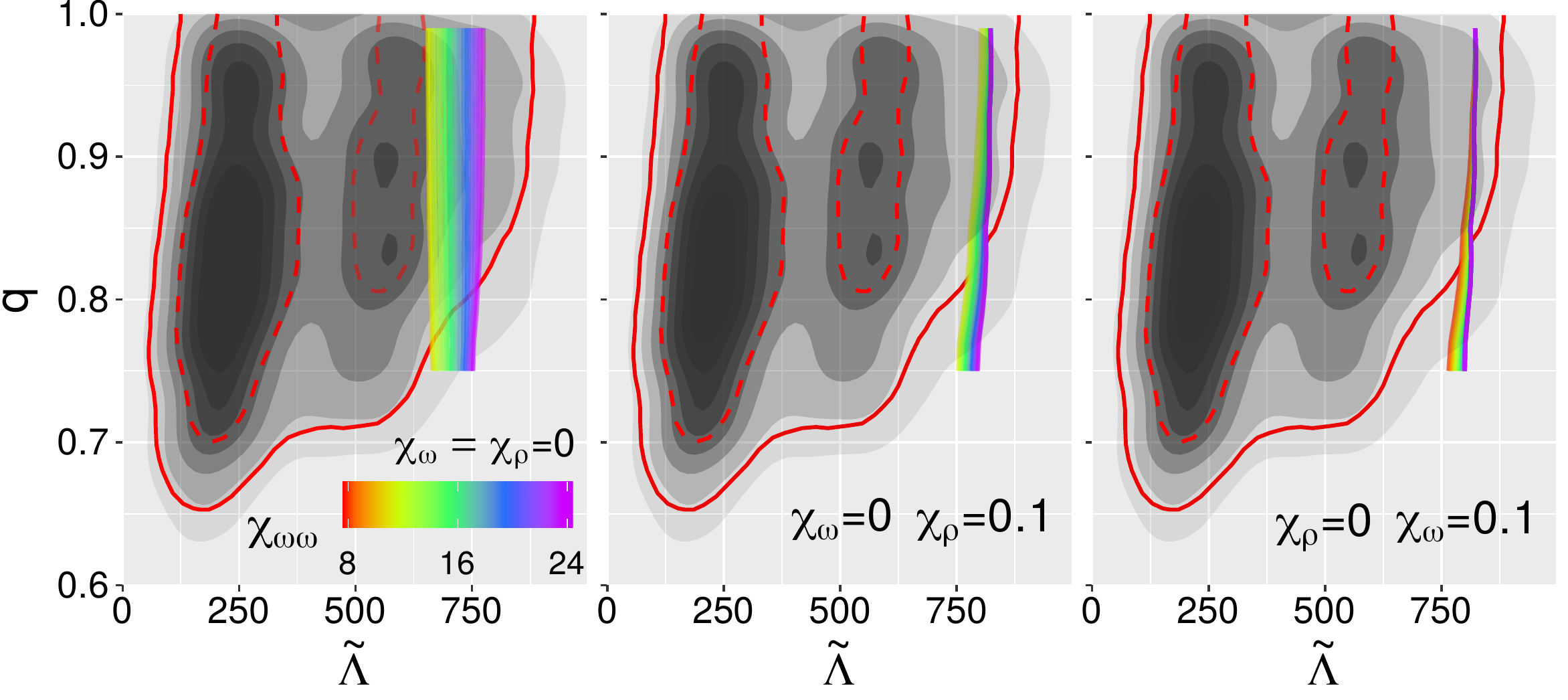}
\end{tabular}
	\caption{$q(\tilde{\Lambda})$ diagrams for binary systems, with $M_{\text{chirp}}=1.186M_{\odot}$, as a function of $\chi_{\omega\omega}$ (color scale) for three sets $(\chi_{\omega}, \chi_{\rho})$: $(0,0)$ (left), $(0.1,0)$ (center), and $(0,0.1)$ (right). 
    The probability distribution function $P(q,\tilde{\Lambda})$ from the LIGO/Virgo collaboration \cite{LIGOScientific:2018hze} is displayed in  gray gradient, while  the red dashed and solid 
	lines indicate the 50\% and 90\% credible regions, respectively.}
	\label{fig:lambda}
	\end{center}
\end{figure}

 We have determined the binary effective tidal deformability $\tilde{\Lambda}$ for $M_{\text{chirp}}=1.186M_{\odot}$ and
$0.73\leq q=M_1/M_2 \leq1$ (GW170817 event \cite{LIGOScientific:2018cki}).
Figure \ref{fig:lambda} shows the $q(\tilde{\Lambda})$ diagram as function of $\chi_{\omega\omega}$ 
(color scale) for three $\{\chi_{\omega}, \chi_{\rho}\}$ sets.
The hybrid EoS are concentrated around  $\tilde{\Lambda}=650-750$ and fall inside the 90\% credible region for $0.8\leq q\leq1$ when $(\chi_{\omega}=0, \chi_{\rho}=0)$. Highly asymmetric systems, $q<0.8$, have a purely hadronic NS component with low $M$ and large $R$, giving rise to larger $\tilde{\Lambda}$. 
For the  other two sets,  $(\chi_{\omega}=0.1, \chi_{\rho}=0)$ and $(\chi_{\omega}=0, \chi_{\rho}=0.1)$, (middle and right panels)  $\tilde{\Lambda} \sim 800$, and, therefore, they are still  compatible with the GW170817 event if moderate asymmetries are considered, i.e., $q>0.8$.

\begin{figure}[!t]
\begin{tabular}{c}
\includegraphics[width=1.\linewidth]{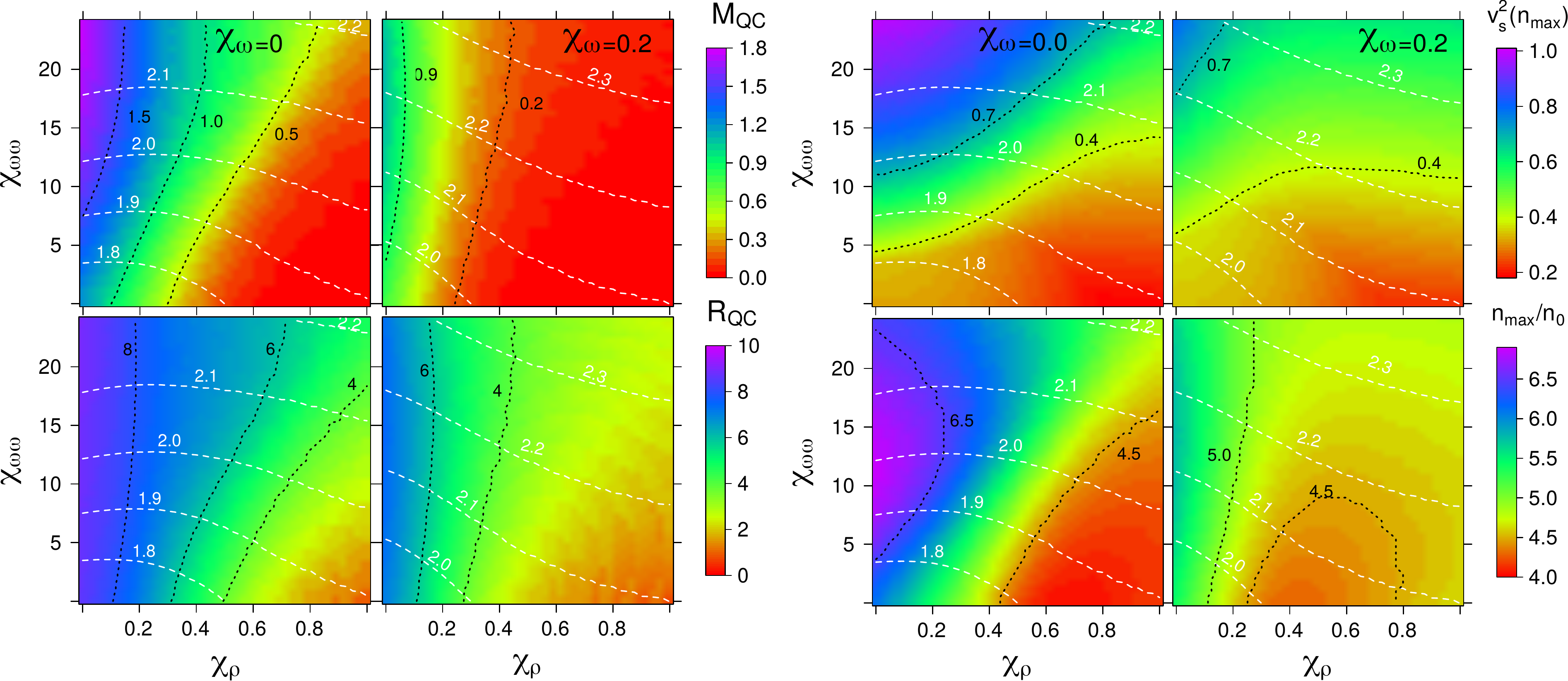}
\end{tabular}
	\caption{The quark core mass $M_{\text{QC}}$ [$M_{\odot}$] (top left),  the quark core radius $R_{\text{QC}}$ [km] (bottom left), the squared speed of sound at the central density [$c^2$] (top right),  and central densities [$n_0$] of $M_{\text{max}}$ (bottom right) as a function of $(\chi_{\omega\omega},\chi_{\rho})$. 
The white dashed and black dotted lines represent, respectively,  the value of $M_{\text{max}}$ [$M_{\odot}$] and specific $\{M_{\text{QC}},R_{\text{QC}}, v_s^2(n_{max}), n_{max}/n_0\}$ contour lines. }
	\label{fig:ns}
\end{figure}

Figure \ref{fig:ns} shows the effect of $\{ \chi_{\omega \omega},\,\chi_{\rho},\, \chi_{\omega} \}$ on various NS properties: the quark core mass $M_{\text{QC}}$, the quark core radius $R_{\text{QC}}$, the speed of sound at the central density $v_s^2(n_{\text{max}})$, 
and the central density $n_{\text{max}}$. Several conclusions can be drawn:
i) while $M_{\text{QC}}$ increases with an increase of $\chi_{\omega \omega}$, the opposite effect occurs for
$\chi_{\rho}$ or/and $\chi_{\omega}$, i.e. for larger values of these last two parameters the mass of the quark core reduces a lot. As a result, large quark cores require
small values of  $\chi_{\rho}$ and $\chi_{\omega}$ and a large
$\chi_{\omega \omega}$ coupling; ii) considering $\chi_{\omega}=\chi_{\rho}=0$, massive NS are only realized for $\chi_{\omega \omega}\gtrsim 12$, with very large radii $R_{\text{QC}}$ and 
speed of sound values $v_s^2(n_{\text{max}})$ close to 1; iii)  $v_s^2$ increases quite rapidly with $\chi_{\omega \omega}$ but decreases with  $\chi_{\rho}$; iv) the largest central densities occur for $\chi_{\omega}=\chi_{\rho}=0$; v) it is possible to have a quark core having a mass of the order of 1/3 of the total star mass, and a speed of sound  $v_s^2(n_{\text{max}})\sim 0.6$. The  softening of the EoS with a finite $\chi_\rho$ due to the earlier onset of strange quarks explains this behavior.

\begin{figure}[!t]
\begin{center}
\begin{tabular}{c}
\includegraphics[width=0.6\linewidth]{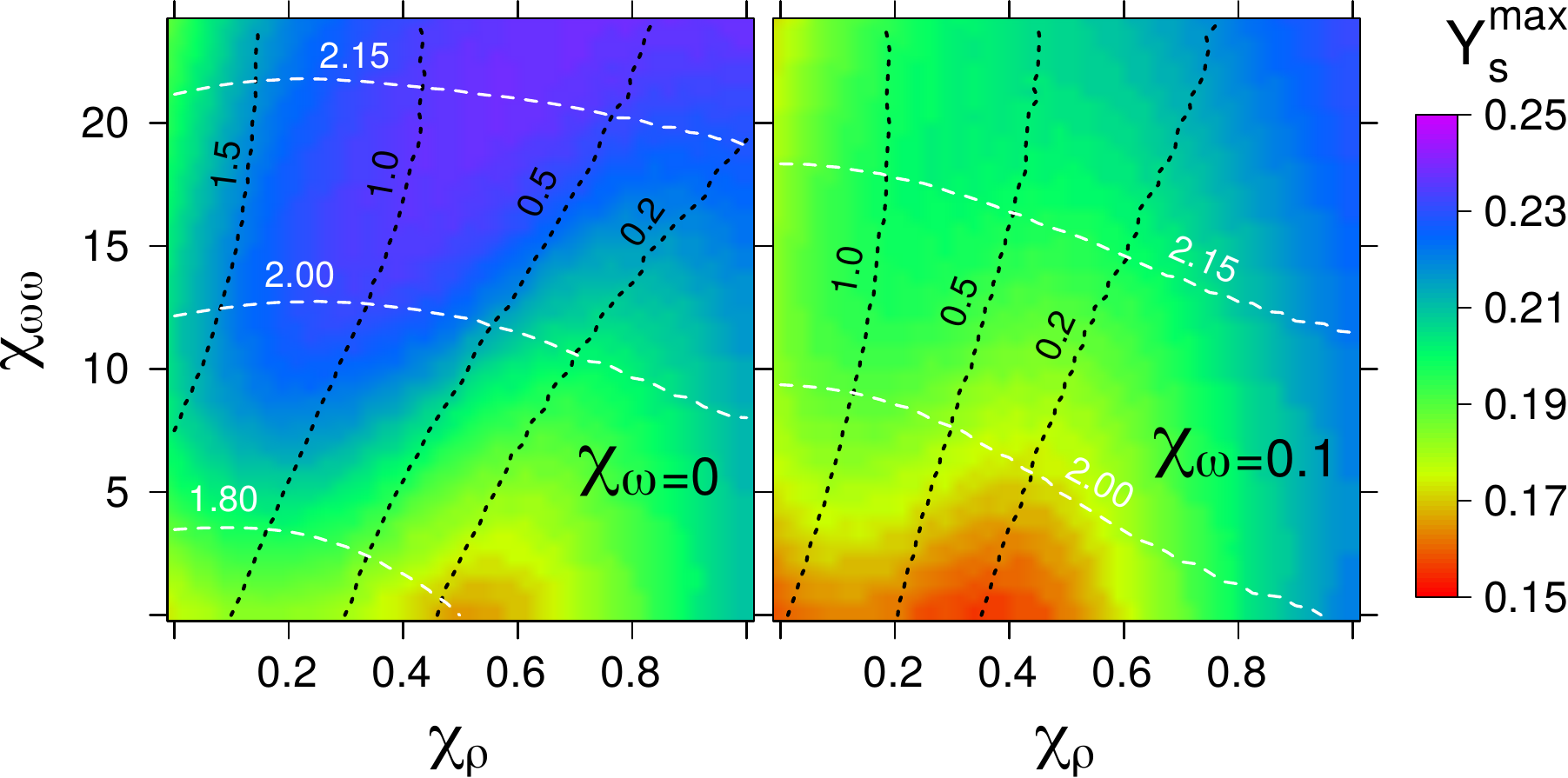}
\end{tabular}
	\caption{The fraction of $s$-quarks,
          $Y_{s}^{\text{max}}$, at the center of the maximum mass star, as a function of $\chi_{\omega\omega}$	and $\chi_{\rho}$ for $\chi_{\omega}=0$ (left) and $0.1$ (right). 
The white dashed and black dotted lines represent, respectively,  the value of the star maximum mass $M_{\text{max}}$ [$M_{\odot}$] and specific $M_{\text{QC}}$ [$M_{\odot}$] contour lines.}
	\label{fig:ys}
\end{center}
\end{figure}

The presence of strangeness inside the star can be understood considering Fig. \ref{fig:ys}, where the  strangeness fraction at the center of the maximum mass star is plotted as a function of $\chi_{\omega \omega}$ and $\chi_{\rho}$ for two values of $\chi_{\omega}$, 0 and 0.1. The largest fraction of strangeness is attained for $\chi_{\omega}=0$, $\chi_{\rho}=0.5$ and $\chi_{\omega \omega}\approx 20$. It is interesting to notice that under these conditions the speed of sound squared is of the order of 0.7 and the strangeness fraction of the order of 0.25. Taking   $\chi_{\rho}=0$ the speed of sound square takes values close to 1 and the strangeness fractions drops to values of the order of 0.18. 
The presence of a larger amount of $s$-quarks prevents the speed of sound to take values close to one, but also reduces the mass of the quark core.

We have studied under which conditions two solar mass hybrid stars may exist. It was shown that a phase transition into a quark phase requires a harder isoscalar channel for the nucleonic EoS, than expected from just considering nucleonic degrees of freedom. 
We have shown that the present constraints on mass, radius, and tidal
deformability from NS observations allow for the existence of hybrid stars with  large quark cores. The $\chi_{\omega \omega}$ channel is important in generating massive quark cores,
even at moderate quark masses, e.g., $1.4M_{\odot}$ stars,  
with  $\tilde{\Lambda}<800$. However, The $\chi_{\omega \omega}$ channel may originate very large values for the speed of sound.  On the other hand, the $\chi_{\rho}$ channel controls the amount of strangeness inside the star, and reduces the speed of sound.

\end{document}